\documentstyle[12pt,a4wide]{article}
\newcommand{\be}{\begin{equation}}
\newcommand{\ee}{\end{equation}}
\newcommand{\bea}{\begin{eqnarray}}
\newcommand{\eea}{\end{eqnarray}}
\newcommand{\bean}{\begin{eqnarray*}}

\newcommand{\eean}{\end{eqnarray*}}
\font\upright=cmu10 scaled\magstep1 \font\sans=cmss10
\newcommand{\ssf}{\sans}
\newcommand{\stroke}{\vrule height8pt width0.4pt depth-0.1pt}
\newcommand{\Z}{\hbox{\upright\rlap{\ssf Z}\kern 2.7pt {\ssf Z}}}

\newcommand{\C}{{\rlap{\rlap{C}\kern 3.8pt\stroke}\phantom{C}}}
\newcommand{\R}{\hbox{\upright\rlap{I}\kern 1.7pt R}}
\newcommand{\CP}{\C{\upright\rlap{I}\kern 1.5pt P}}
\newcommand{\PP}{\hbox{\upright\rlap{I}\kern 1.5pt P}}

\newcommand{\identity}{{\upright\rlap{1}\kern 2.0pt 1}}

\newcommand{\HH}{\mbox{\hbox{\upright\rlap{I}\kern 1.7pt H}}}

\newcommand{\zb}{{\bar z}}

\newcommand{\fr}{\frac}

\newcommand{\ra}{\rightarrow}
\newcommand{\al}{\alpha}
\newcommand{\bt}{\beta}
\newcommand{\pr}{\partial}
\newcommand{\hs}{\hspace{5mm}}

\newcommand{\dg}{\dagger}

\newcommand{\acc}{\\[3mm]}

\input{epsf}
\begin{document}
\begin{center}{\bf
Axially symmetric  $SU(3)$ Gravitating Skyrmions}\\
\vglue 0.5cm { Theodora Ioannidou$^\dagger${\footnote{{\it Email}:
ti3@auth.gr}},
Burkhard Kleihaus $^\ddagger$
{\footnote{{\it Email}: kleihaus@theorie.physik.uni-oldenburg.de}}
and 
Wojtek Zakrzewski$^*$
{\footnote{{\it Email}: W.J.Zakrzewski@durham.ac.uk}}
} \\
\vglue 0.3cm
$^\dagger${\it Maths Division, School of Technology, University of
Thessaloniki, Thessaloniki 54124,  Greece}\\
$^\ddagger${\it Institut f\"ur Physik, Universit\"at Oldenburg, Postfach 2503
D-26111 Oldenburg, Germany\\
$^*${\it Department of Mathematical Sciences, University
of Durham, Durham DH1 3LE, UK}\\}
\end{center}

Axially symmetric  gravitating multi-skyrmion  configurations are obtained 
using the   harmonic map ansatz introduced  in \cite{IPZ1}.
In particular, the effect of gravity on the energy and baryon 
densities of the $SU(3)$ non-gravitating multi-skyrmion configurations is
studied in detail.

\section{Introduction}

The Einstein-Skyrme model can be thought of as a nonlinear field theory
which describes the interaction between a skyrmion (ie. a baryon) and a
black hole.
So far, mainly  spherically symmetric $SU(N)$ gravitating skyrmions and
black hole configurations with Skyrme hair have been obtained
\cite{DHS}-\cite{BHIZ}.
Recently, in \cite{S1,S2} $SU(2)$ static axially symmetric regular
 and black hole
solutions have been derived numerically indicating
that the energy densities of such system depend on the
coupling  constant. In particular, as the coupling constant increases
the energy (and baryon) density becomes either denser for the gravitating
 skyrmions or sparser for the black hole solutions.
In what follows we will construct the $SU(3)$ field
configurations approximating axially symmetric gravitating
skyrmions and black holes (i.e. close to the solutions
of the full equations) and investigate their properties.

The $SU(N)$ Einstein-Skyrme action reads:
 \be
  S=\int \left[\fr{R}{16\pi G}+\fr{\kappa^2}{4} \mbox{tr}\left(K_{\mu}\,K^\mu 
\right) +\fr{1}{32e^2}\mbox{tr}\left(\left[K_\mu,K_\nu\right]\left[K^\mu,
K^\nu\right
]\right)\right]
\sqrt{-g}\, d^4x
  \label{ac}
  \ee
where $K_\mu=\pr_\mu U U^{-1}$ for $\mu=0,1,2,3$;  $U$ is the  $SU(N)$ 
chiral field; $g$ denotes the determinant of the
  metric; while $\kappa$, $e$ are coupling constants and
$G$ represents Newton's constant.
In order for the finite-energy configurations to exist the Skyrme field
has to go to a constant matrix at spatial infinity:  $U\ra I$ as
$|x^\mu|\ra \infty$.
This effectively compactifies the three-dimensional Euclidean space
into $S^3$ and hence implies that the field configurations of the Skyrme
 model can be considered as maps from $S^3$ into $SU(N)$.

The variation of the action (\ref{ac}) with respect to the metric 
$g^{\mu \nu}$ leads to the Einstein equations
\be
R_{\mu \nu}-\fr{1}{2}g_{\mu \nu}R=8\pi G T_{\mu \nu}
\label{E}
\ee
with the stress-energy tensor   given by
\be
T_{\mu\nu}\!\!=-\fr{\kappa^2}{2}\mbox{tr}\!\left(K_\mu K_\nu
-\fr{1}{2}g_{\mu \nu}
K_\al K^\al \right)- \fr{1}{8e^2} 
\mbox{tr}\!\left(g^{\al \bt}\left[K_\mu,K_\al\right]\left[K_\nu,K_\bt\right]-
\fr{1}{4}g_{\mu \nu} \left[K_\al,K_\bt\right]\left[K^\al,K^\bt\right]\right).
\label{T}
\ee

The starting point of our investigation is the introduction of the
coordinates $r,z,\zb$ on $\R^3$. In terms of the usual spherical
coordinates $r,\theta,\phi$ the Riemann sphere variable $z$ is given by
$z=e^{i\phi} \tan(\theta/2)$. In this system of coordinates the
ansatz for the static axially symmetric isotropic metric reads
\begin{equation}
ds^2=-fdt^2+\fr{m}{f}dr^2+
\fr{r^2}{f}\left(\fr{(m-l)\bar{z}}{z(1+|z|^2)^2}dz^2
+\fr{z(m-l)}{\bar{z}(1+|z|^2)^2}d\bar{z}^2
+\frac{2(m+l)}{(1+|z|^2)^2}
\,dz d\bar{z} \right)
 \label{s}
\end{equation}
where the square-root of the determinant of the metric is of the form
\begin{equation}
\sqrt{-g}=\frac{i\,m\sqrt{l}}{f}  \,\frac{2r^2}{(1+|z|^2)^2}
\end{equation}
while  the metric profiles $f$, $l$, $m$ are functions of 
$(r,|z|^2)$.

The Einstein-Skyrme system has a topological current which is covariantly 
conserved, yielding the topological charge \cite{GKK}
\be
B=\int \sqrt{-g} \,B^0\, d^3x
\label{b}
\ee
where 
\be
B^\mu=-\fr{1}{24\pi^2\sqrt{-g}}\,\varepsilon^{\mu\nu\al\bt}\,\mbox{tr}\left
(K_\nu K_\al K_\bt\right)
\ee
and  $\varepsilon^{\mu\nu\al\bt}$ is the (constant) fully antisymmetric
tensor.

In the stereographically projected coordinates the  action (\ref{ac}) for 
the  metric (\ref{s}) becomes equal to
\bea
 S\!\!\!\!&=&\!\!\!\!\!\!\int dt\, dr \,dz \,d\bar{z}\,  \sqrt{-g}\,
\bigg[\fr{R}{16\pi G}+\fr{\kappa^2}{4} \mbox{tr}\left(
g^{rr}K_r^2+g^{zz}K_z^2+g^{\bar{z}\bar{z}}K_{\bar{z}}^2+2g^{z\bar{z}}|K_z|^2
\right) \nonumber\\
&&\!\!\!\!\!\!\!+\fr{1}{16e^2}\mbox{tr}\left(g^{rr}g^{zz}
\left[K_r,K_z \right]^2\!+
g^{rr}g^{\bar{z}\bar{z}} \left[K_r,K_{\bar{z}}\right]^2\!+
\left(g^{\bar{z}\bar{z}}g^{zz}-g^{{z\bar{z}}^2}\right)
\left[K_z,K_{\bar{z}}\right]^2\!
+2g^{rr}g^{z\bar{z}}\bigg|\!\left[K_r,K_z\right]\bigg|^2\right)\bigg]
\nonumber\\
  \label{ac1}
  \eea
while the baryon number takes the form
\be
B=-\fr{1}{8\pi^2}\int\mbox{tr}\left(K_r\left[K_z,K_{\bar{z}}\right
]\right) dr\, dz\,d\bar{z}.
\label{B}
\ee

Note that, the metric (\ref{s}) leads to five nontrivial 
equations for the three functions $f$, $m$ and $l$. In what follows, 
we consider static fields only (i.e. time independent Skyrme fields).
Following  the approach of  \cite{KK} we consider 
linear superpositions of the stress-energy tensor
\be
{\cal M}_1=T^r_r+T^\theta_\theta+T^\phi_\phi-T^0_0,\hs
{\cal M}_2=g_{rr}\left(T^r_r+T^\phi_\phi\right),\hs
{\cal M}_3=g_{rr}\left(T^r_r+T^\theta_\theta\right). 
\label{ten}
\ee
Then, the Einstein equations (\ref{E}) take the form
\bea
8\pi G{\cal M}_1\!\!\!\!
& = &\!\!\!\!
\fr{f_{rr}}{f}+\fr{2}{r}\fr{f_r}{f}-\fr{f_r^2}{f^2}+\fr{1}{2}
\fr{f_r}{f}\fr{l_r}{l}+\fr{(1+|z|^2)^2}{r^2} \left[|z|^2 \left(\fr{f''}{f}
-\fr{f'^2}{f^2}\right)+\fr{1}{2}
\fr{f'}{f}\left(1+|z|^2\fr{l'}{l}\right)\right]\nonumber\acc
8\pi G{\cal M}_2\!\!\!\!
& = &\!\!\!\!\fr{1}{2}\fr{m_{rr}}{m}+\fr{1}{r}\fr{m_r}{m}
-\fr{1}{2}\fr{m_r^2}{m^2}+\fr{(3|z|^2-1)|z|^2}{4r^2}
\fr{m'}{m}+\fr{(1+|z|^2)^2|z|^2}{2r^2}\left(\fr{m''}{m}-\fr{m'^2}{m^2}\right)
\nonumber\\
&+&\fr{1}{2r}\fr{l_r}{l}+
\fr{(1+|z|^2)^2 |z|^2}{4r^2}\left(2\fr{l''}{l}-\fr{l'^2}{l^2}\right)
+\fr{(1+|z|^2)}{2r^2}\fr{l'}{l}\nonumber\\
&+&\fr{1}{4}\fr{m_r}{m} \fr{l_r}{l}+\fr{(1+|z|^2)^2 |z|^2}{4r^2}\left(
2\fr{f'^2}{f^2}-\fr{m'}{m}\fr{l'}{l}\right)\nonumber\acc
8\pi G{\cal M}_3\!\!\!\!
& = &\!\!\!\!
\fr{1}{2}\fr{l_{rr}}{l}+\fr{3}{2r}\fr{l_r}{l}-\fr{1}{4}\fr{l_r^2}{l^2}
+\fr{(1+|z|^2)}{2r^2}\fr{l'}{l}+\fr{(1+|z|^2)^2 |z|^2}{2r^2}\left(\fr{l''}{l}
-\fr{1}{2}\fr{l'^2}{l}\right),\label{AE}
\eea
where $l_r$ denotes the partial derivative of the function $l$ with respect to $r$ 
and $l'$ the partial derivative with respect to the argument $|z|^2$ 
(similarly for other functions). Also, the expressions in (\ref{ten}) 
simplify to
\bea
{\cal M}_1\!\!\!\!&=&\!\!\!\!-\fr{1}{16e^2}\fr{f^2\left(l+m\right)}{m^2\,l}\,
\fr{(1+|z|^2)^2}{r^2}
\mbox{tr}\bigg|[K_r,K_z]\bigg|^2
+\fr{1}{32e^2}\fr{f^2}{ml}\,\fr{(1+|z|^2)^4}{r^4}\,
\mbox{tr}[K_z,K_{\bar{z}}]^2\nonumber\acc
{\cal M}_2\!\!\!\!&=&\!\!\!\!
\fr{\kappa^2}{8}\fr{m}{f}\fr{(1+|z|^2)^2}{|z|^2r^2}\,
 \mbox{tr} \left(z^2K_z^2+\bar{z}^2K_{\bar{z}}^2+2|z|^2|K_z|^2\right)\nonumber\\
&&\!\!\!\!+\fr{1}{32e^2}\fr{f}{l}\fr{(1+|z|^2)^2}{|z|^2\,r^2} \,\mbox{tr}
\left(z^2[K_r,K_z]^2+\bar{z}^2[K_r,K_{\bar{z}}]^2-2|z|^2\bigg|[K_r,K_z]
\bigg|^2\right)\nonumber\acc
{\cal M}_3\!\!\!\!&=&\!\!\!\!-\fr{\kappa^2}{8}\fr{m}{l}\fr{(1+|z|^2)^2}
{|z|^2r^2}\,
 \mbox{tr} \left(z^2K_z^2+\bar{z}^2K_{\bar{z}}^2-2|z|^2|K_z|^2\right)
\nonumber\\
&&\!\!\!\!-\fr{1}{32e^2}\fr{f}{m}\fr{(1+|z|^2)^2}{|z|^2\,r^2} \,\mbox{tr}
\left(z^2[K_r,K_z]^2+\bar{z}^2[K_r,K_{\bar{z}}]^2+2|z|^2\bigg|[K_r,K_z]
\bigg|^2\right).
\label{M}
\eea
Next we consider the static Einstein-Skyrme equations 
and  construct field configurations approximating 
their axially symmetric solutions based on the 
harmonic map ansatz using the formulations of \cite{IPZ1}.

\section{The Harmonic Map Ansatz}

The idea of the harmonic map ansatz (i.e. the generalisation
of the rational map ansatz of Houghton et al \cite{HMS}) 
involves the separation of the radial and angular 
dependence of the fields  \cite{IPZ1} as
\be
U=e^{2ih(r)\left(P-I/N \right)}
=e^{-2ih(r)/N}\left[I+\left(e^{2ih(r)}-1\right)P\right]
\label{U}
\ee
where $P$ is a $N\times N$ hermitian projector which depends only on the
angular variables $(z,\bar{z})$ and $h(r)$ is the corresponding 
profile function.
Note that the matrix $P$ can be thought of as a mapping from $S^2$ into 
$CP^{N-1}$. Thus, $P$ can be written as
\be
P(V)=\fr{V \otimes V^\dg}{|V|^2},
\label{for}
\ee
where $V$ is a $N$-component complex vector (dependent on $z$ and
$\bar{z}$). 
Following \cite{DinZak}, we define a new operator $P_+$,  by its action
on any vector $v \in \C^N$, as
\be
P_+ v=\pr_\xi v- \fr{v \,(v^\dg \,\pr_\xi v)}{|v|^2},
\ee
and then define further vectors $P^k_+ v$ by induction: 
$P^k_+ v=P_{+}(P^{k-1}_+ v)$.
As shown in \cite{Za}, these vectors have many interesting properties
when $V$ is holomorphic. These properties either follow directly from
 the definition of the $P_+$  operator or are very easy to prove and they 
lead to many simplifications  of the expressions 
for the action  and the energy-momentum tensor.

For (\ref{U}) to be well-defined at the origin, 
the radial profile function $h(r)$ has to satisfy
$h(0)=\pi$ while the boundary value $U \ra I$ at $r=\infty$ requires that 
$h(\infty)=0$.
As shown in \cite{IPZ1},
an attractive feature of the ansatz (\ref{U}) is that it leads to a
simple expression for the energy density which can  be successively minimized
with respect to the parameters of the projector $P$ and then with respect
to the shape of the profile function $h(r)$. 
This procedure gives good approximations to multi-skyrmion field 
configurations in flat space \cite{IPZ1}.

\begin{figure}
\centering
\epsfysize=9cm
\mbox{\epsffile{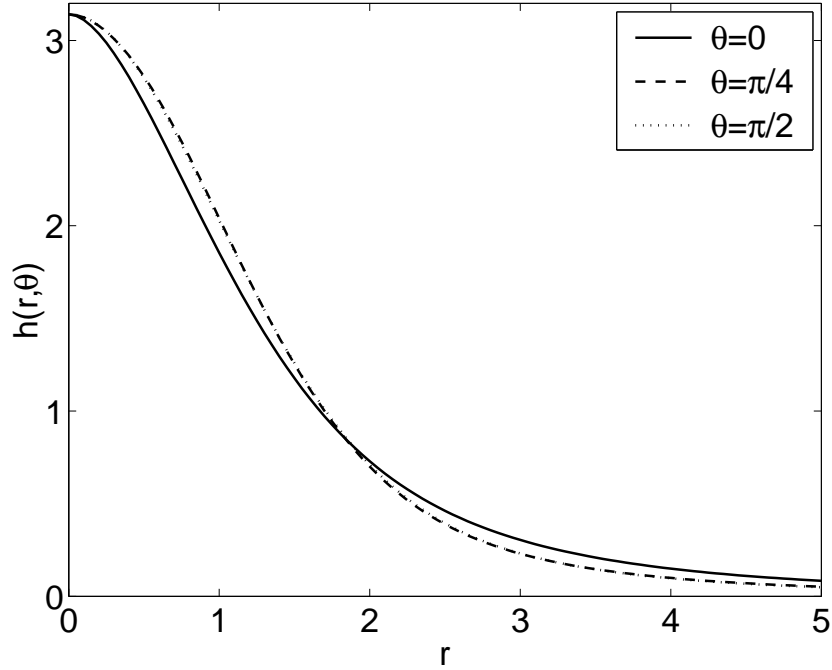}}
\caption{\label{sprof} The $|z|^2$ dependence of the Skyrme field profile 
function. [Recall, $\vert z\vert =\tan{\theta\over2}$].}
\centering
\end{figure}

Since we are interested in axially symmetric field configurations
in curved space the action possesses, in addition to the 
explicit $|z|^2$ dependence due to the complex vector $v$, an 
implicit $|z|^2$ dependence
induced by the metric functions. Consequently, we cannot simply integrate
the action over $|z|^2$ (as in flat space).
Therefore, we consider a slightly modified ansatz by allowing the profile 
function $h$ to depend on  $r$ and $|z|^2$ (shown in Figure \ref{sprof}). 
In fact, this ansatz
(concerned with  the Skyrme part) involves a generalisation of the rational 
map ansatz, the so-called improved harmonic map ansatz introduced in 
 \cite{us}.
Indeed \cite{us} presents better approximations (i.e. with  lower energies)
 of pure  $SU(2)$ skyrmions  and so are better approximants of the
exact solutions obtained numerically by solving the full equations.

The action (\ref{ac1}) after the substitution of (\ref{U}) with $h=h(r,|z|^2)$
and having used the properties of the harmonic maps becomes
\bea
S\!\!\!\!&=&\!\!\!\!\int\bigg({\cal K} \,r^2\,\fr{R}{8\pi G}
-\kappa^2\,A_N\, r^2\,h_r^2
-\kappa^2 B_N \,h'^2
-\left(\kappa^2\,{\cal N}_1+\fr{1}{e^2}
 {\cal N}_2 h_r^2+\fr{1}{e^2}\,{\cal N}_3\, \fr{h'^2}{r^2}\right)
\sin^2 h
\nonumber\\
&&-\fr{1}{e^2}\, {\cal I}\,\fr{\sin^4 h}{r^2}\bigg)\, dt\,dr\,dz\, d\bar{z}
\label{s1}
\eea
where
\bea
{\cal K}&=&i  \fr{m\sqrt{l}}{f}\,\fr{1}{(1+|z|^2)^2}\nonumber\\
A_N&=&2i\,\fr{N-1}{N} \fr{\sqrt{l}}{(1+|z|^2)^2}\nonumber\\
B_N&=&2i\,\fr{N-1}{N}\, \sqrt{l}\, |z|^2\nonumber\\
{\cal N}_1&=&i\, \fr{(m+l)}{\sqrt{l}}\,\fr{|P_{+}V|^2}{|V|^2}\nonumber\\
{\cal N}_2&=&i\, \fr{f}{m} \,\fr{(m+l)}{\sqrt{l}}\,\fr{|P_{+}V|^2}{|V|^2}\, 
\nonumber\\
{\cal N}_3&=&i\, \fr{f}{\sqrt{l}} 
\,\fr{|P_{+}V|^2}{|V|^2}\, (1+|z|^2)^2\,|z|^2
\nonumber\\
{\cal I}&=&i\,\fr{f}{\sqrt{l}}\,\fr{|P_+V|^4}{|V|^4}\, (1+|z|^2)^2 
\eea
The baryon number (\ref{B}) 
coincides with the expression for the topological  charge of the $CP^{N-1}$ 
sigma model (up to an overall profile dependent factor) since
\bea
B&=&\fr{i}{\pi^2}\int\mbox{tr}\left(P\left[P_z,P_{\bar{z}}\right]\right)
\,dz\,d\bar{z}
\int_{0}^{\infty} \sin^2 h \,h_r\,dr
\nonumber\\
&=&\fr{i}{2\pi}\int\,\fr{|P_+V|^2}{|V|^2}\, dz\,d\bar{z}
\label{b1}
\eea
The $r$ integration can be performed 
although $h=h(r,\vert z\vert\sp2)$  since 
$\int_0\sp{\infty} \sin\sp2h\,h_r\,dr=h(0)/2$ and we have 
required that $h(0,\vert z\vert\sp2)=\pi.$

The variation of (\ref{s1}) with respect to  $h$  gives its 
equation of motion:
\bea
\pr_x\left[\left(A_N\,x^2+{\cal N}_2\,
\sin^2 h\right) h_x\right]+
\left[\left(B_N+\fr{1}{x^2}\,{\cal N}_3
\sin^2 h\right) h'\right]'&&\nonumber\\
-\left(\fr{1}{2}\,{\cal N}_1
+\fr{1}{2}\,{\cal N}_2\,h_x^2+\fr{1}{2}\,{\cal N}_3\,\fr{h'^2}{x^2}
+{\cal I}\,\fr{\sin^2 h}{x^2}\right)
\sin(2h)\!\!\!&=&\!\!\!0 
\label{ec}
\eea
where the dimensionless coordinate $x=e\kappa r$ has been  introduced.

In addition,  the left-hand side components (\ref{M}) of the Einstein equations 
 (\ref{AE}) become
\bea
{\cal M}_1\!\!\!\!&=&\!\!\!\!e^2\kappa^4\fr{f^2}{ml}\,\fr{\sin^2 h}{x^2}
 \fr{|P_+ V|^2}{|V|^2} (1+|z|^2)^2\left[
\fr{l+m}{m}\,h_x^2+\fr{(1+|z|^2)^2}{x^2}\left( |z|^2\,h'^2
+\sin^2 h\,\fr{|P_+ V|^2}{|V|^2}
\right)\right]\nonumber\hs\\
{\cal M}_2\!\!\!\!&=&\!\!\!\!
e^2\kappa^4(1+|z|^2)^2\left[
-2\, |z|^2\,
\fr{N-1}{N} \,\fr{h'^2}{x^2}+
\left(-1+\fr{f}{l} 
\,h_x^2\right)\,\fr{\sin^2 h}{x^2}
\, \fr{|P_+ V|^2}{|V|^2}\right]\nonumber\\
{\cal M}_3\!\!\!\!&=&\!\!\!\!
e^2\kappa^4\left(-\fr{m}{l}+\fr{f}{m} \,h_x^2\right)
\fr{|P_+ V|^2}{|V|^2}\,(1+|z|^2)^2 \, \fr{\sin^2 h}{x^2}.
\eea
In \cite{HMS,IPZ1} the explicit forms of  
the holomorphic vector $V$
 have been obtained by minimizing  the $SU(2)$ and $SU(3)$ 
 skyrmion energy densities, respectively.
In particular,
 the low energy field configurations of the  $SU(2)$  Skyrme model 
with baryon number from one up to seventeen and of the
$SU(3)$  Skyrme model 
with baryon number from one
up to six, which are not the  $SU(2)$ embeddings, have been constructed. 
From these, the axisymmetric field configurations are the following:

\begin{itemize}
\item  {\it $SU(2)$ $B=2$ Skyrmion}: 
The associated vector $V$ has the simple form
\cite{HMS}
\be
V=(z^2,1)^t;\hs \hs \hs
\fr{|P_+ V|^2}{|V|^2}=\fr{4 |z|^2}{(1+|z|^4)^2}
\label{vsu2b2}
\ee
Recently, this case has been studied in \cite{S1}.

\item{\it $SU(3)$ $B=4$ Skyrmion}:
The harmonic vector $V$ is of the form \cite{IPZ1}:
\be
V=\left(z^4, a\,z^2,\, 1 \right)^t,\hs \hs \hs
\fr{|P_+V|^2}{|V|^2}=\fr{4|z|^2\left(a^2(1+|z|^8)
+4|z|^4\right)}{\left(|z|^8+a^2|z|^4+1\right)^2}
\label{fB4}
\ee
for $a=2.7191$.
The corresponding configuration has the shape of
{\it two tori  on top of each other close to the equator of the 
sphere} and
its energy and baryon density  are invariant under a rotation around 
the $z$-axis,  which is
equivalent to the $SU(3)$ transformation:  $U \rightarrow A^{-1} U A$
for  $A = \mbox{diag}(e^{-2\,i\alpha},\, 1, e^{-2 i\alpha})$. 
\end{itemize}
Next we use the vectors $V$ given by (\ref{vsu2b2}) and 
 (\ref{fB4}) in order to obtain approximations  to the gravitating 
$SU(2)$ $B=2$ and $SU(3)$ $B=4$ skyrmions, respectively.

\section{Numerical Solutions}

Due to convenience of our numerical scheme  we transform our expressions 
back to
 spherical coordinates where the metric is parametrized
in terms of  $(x,\theta,\phi)$ as
\be
ds^2=-f dt^2+\fr{m}{f}\, dx^2+\fr{m}{f}\, x^2 d\theta^2+\fr{l}{f}\, r^2
\sin^2 \theta \,d\phi^2
\ee
while the Einstein equations (\ref{E}) take the form
\bea
2\alpha {M}_1\!\!\!&=&\!\!\!\fr{f_{xx}}{f}+\fr{1}{x^2}\fr{f_{\theta\theta}}{f}
+\fr{2}{x}\fr{f_x}{f}-\fr{1}{x^2}\fr{f_\theta^2}{f^2}-\fr{f_x^2}{f^2}
+\fr{\cot \theta}{x^2} \fr{f_\theta}{f}+\fr{1}{2}\fr{f_x}{f}\fr{l_x}{l}
+\fr{1}{2x^2}\fr{f_\theta}{f}\fr{l_\theta}{l}\hs\hs
\nonumber\acc
2 \alpha {M}_2\!\!\!&=&\!\!\!\fr{1}{2}\fr{m_{xx}}{m}+\fr{1}{x}\fr{m_x}{m}+
\fr{1}{2x^2}\fr{m_{\theta
\theta}}{m}-\fr{1}{2x^2}\fr{m_\theta^2}{m^2}-\fr{1}{2}\fr{m_x^2}{m^2}
+\fr{1}{2x^2}\fr{l_{\theta \theta}}{l}-\fr{1}{4x^2}\fr{l_\theta^2}{l^2}\nonumber
+\fr{1}{2x^2}\fr{f_\theta^2}{f^2}
\\
&&
\!\!\!+\fr{1}{2x}\fr{l_x}{l}+\fr{1}{2x}\fr{m_x}{m}+\fr{1}{4}\fr{m_x}{m}\fr{l_x}{l}
-\fr{1}{4x^2}\fr{m_\theta}{m} \fr{l_\theta}{l}
-\fr{\cot \theta}{2x^2} \fr{m_\theta}{m}+\fr{\cot \theta}{x^2} \fr{l_\theta}{l}
\nonumber\acc
2 \alpha {M}_3\!\!\!&=&\!\!\!\fr{1}{2}\,\fr{l_{xx}}{l}+\fr{1}{2x^2}\,
\fr{l_{\theta\theta}}{l}-\fr{1}{4x^2}\,\fr{l_\theta^2}{l^2}-
\fr{1}{4}\,\fr{l_x^2}{l^2}+\fr{3}{2x}\,
\fr{l_x}{l}+\fr{l_\theta}{l} \cot \theta.
\eea
The dimensionless coupling constants $\al$ and $M_i$ are defined as:
 $\alpha= 4 \pi G \kappa^2$ and 
 ${M}_i ={\cal M}_i/e^2\kappa^4$ for $i=1,2,3$.

This system of coupled non-linear partial differential 
equations has to be solved numerically subject to the given boundary conditions.
To map the infinite interval of the radial variable $r$ onto the 
finite interval $[0,1]$ we introduce the variable $\bar{x}$ as
\be
\bar{x} = \frac{x}{1+x}.
\ee
Since the functions are symmetric under the transformation
$\theta \rightarrow \pi-\theta$, it is sufficient to 
solve the equations for only $0 \leq \theta \leq \pi/2$   
with the boundary conditions 
\be
h_\theta(\bar{x},\theta=\pi/2)
=f_\theta(\bar{x},\theta=\pi/2)
=m_\theta(\bar{x},\theta=\pi/2)
=l_\theta(\bar{x},\theta=\pi/2)=0 \ .
\ee
To satisfy the regularity condition $m(\bar{x},\theta=0)=l(\bar{x},\theta=0)$
we introduce the function 
\be
g(\bar{x},\theta)=\frac{m(\bar{x},\theta)}{l(\bar{x},\theta)}
\ee
and impose the boundary conditions 
$g(\bar{x},\theta=0)=1$, $g(\bar{x}=0,\theta)=1$, $g(\bar{x}=1,\theta)=1$  
and $g_\theta(\bar{x},\theta/\pi/2)=0$.
The numerical computations,  based on the Newton-Raphson method,
 have been performed 
using  the program FIDISOL \cite{fidisol} where 
the partial differential equations are discretized on a non-equidistant grid  
with typical grids sizes $130\times 30$ and estimated relative numerical 
errors  of order $ 10^{-3}$.

It has been observed that, for a fixed  complex vector $V$,
the globally regular solutions of the Einstein-Skyrme model depend
{\it  only} on the 
dimensionless coupling constant $\alpha= 4 \pi G \kappa^2$.
In fact, the numerical simulations show that such field
configurations exist only for a finite range of $\al$: 
 $0\leq \alpha \leq \alpha_{\rm max}$ where
its maximal value  depends on the specific form of  $V$.
For instance,  in 
the SU(2) $B=2$ case where  $V$ is given by  (\ref{vsu2b2}) 
  $\alpha_{\rm max}=0.0338$, 
while in  the SU(3) $B=4$ case where $V$ is given by (\ref{fB4})
 with $a=2.7191$ 
 $\alpha_{\rm max}=0.0209$.
 Assuming that our field configurations are very close to the proper solutions
 of the field equations (as shown in \cite{us}) 
we believe that what we have seen here is also true for the exact  (skyrmion) solutions.

Hence we note that starting from a skyrmion in flat space
a branch of gravitating skyrmions emerges for $\alpha>0$.
This branch  terminates at the maximal value of the 
coupling constant $\alpha_{\rm max}$, where it joints 
smoothly a second branch of solutions, which bends backwards from 
$\alpha=\alpha_{\rm max}$ to $\alpha=0$.

\begin{figure}
\centering
\epsfysize=9cm
\mbox{\epsffile{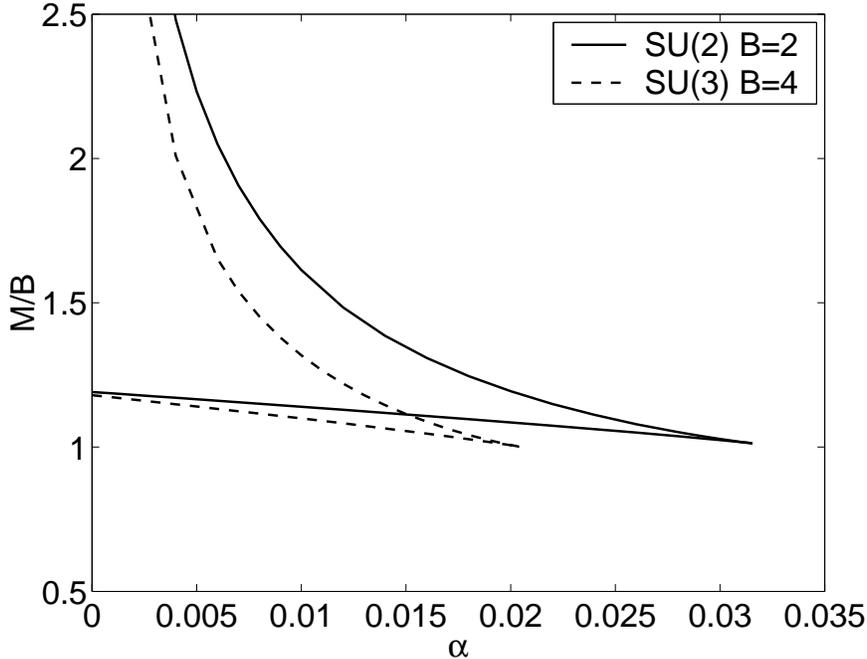}}
\caption{\label{Fig1} The dimensionless mass $M/B$ in terms  of the 
coupling constant $\al$ in the $SU(2)$ and $SU(3)$ case.}
\centering
\end{figure}

Figure \ref{Fig1} presents the dimensionless mass $M/B= -(6 \pi^2 \kappa/e)^{-1}
\left(\int{T_0^0 \sqrt{|g|} d^3r}\right)
/B$
as a function of $\alpha$ 
for the SU(2) $B=2$ and SU(3) $B=4$ skyrmions.
The mass decreases along the first branch with increasing $\alpha$ 
until it reaches its minimum at $\alpha=\alpha_{\rm max}$ 
then it increases along the second branch with decreasing $\alpha$ 
and diverges as $\alpha$ tends to zero.
Accordingly, we refer to the branches with lower (respectively higher) mass 
as the lower (respectively upper) branch.

\begin{figure}
\centering
\epsfysize=9cm
\mbox{\epsffile{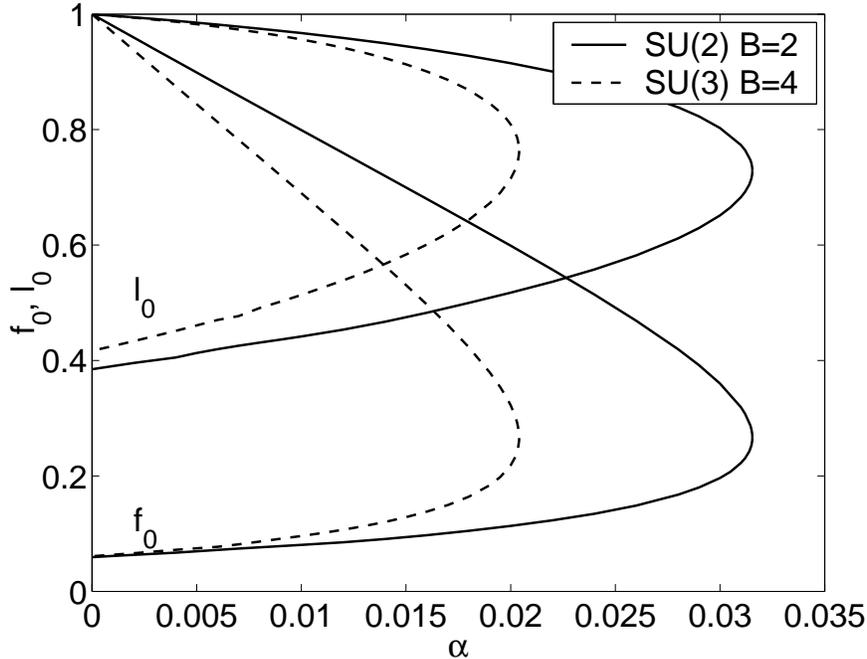}}
\caption{\label{Fig4}The axial part  of the initial metric profiles 
$f(0,\theta)$ and $m(0,\theta)$ as  functions of the gravitational constant
 $\al$ in the $SU(2)$ and $SU(3)$ case.}
\end{figure}

Figure \ref{Fig4} presents the $\al$ dependence of  the metric functions at
the origin  $f_0=f(0,\theta)$ and $l_0=l(0,\theta)$
 for the SU(2) $B=2$ and the SU(3) $B=4$ field configurations.
Note that, $f_0$ and $l_0$ decrease monotonically
for the configurations along the lower branch with increasing $\alpha$ 
and along the upper branch with decreasing $\alpha$; 
they  approach finite values as $\alpha$ tends to zero
on the upper branch.

\begin{figure}
\parbox{\textwidth}
{\centerline{
\mbox{\epsfysize=25.0cm
\epsffile{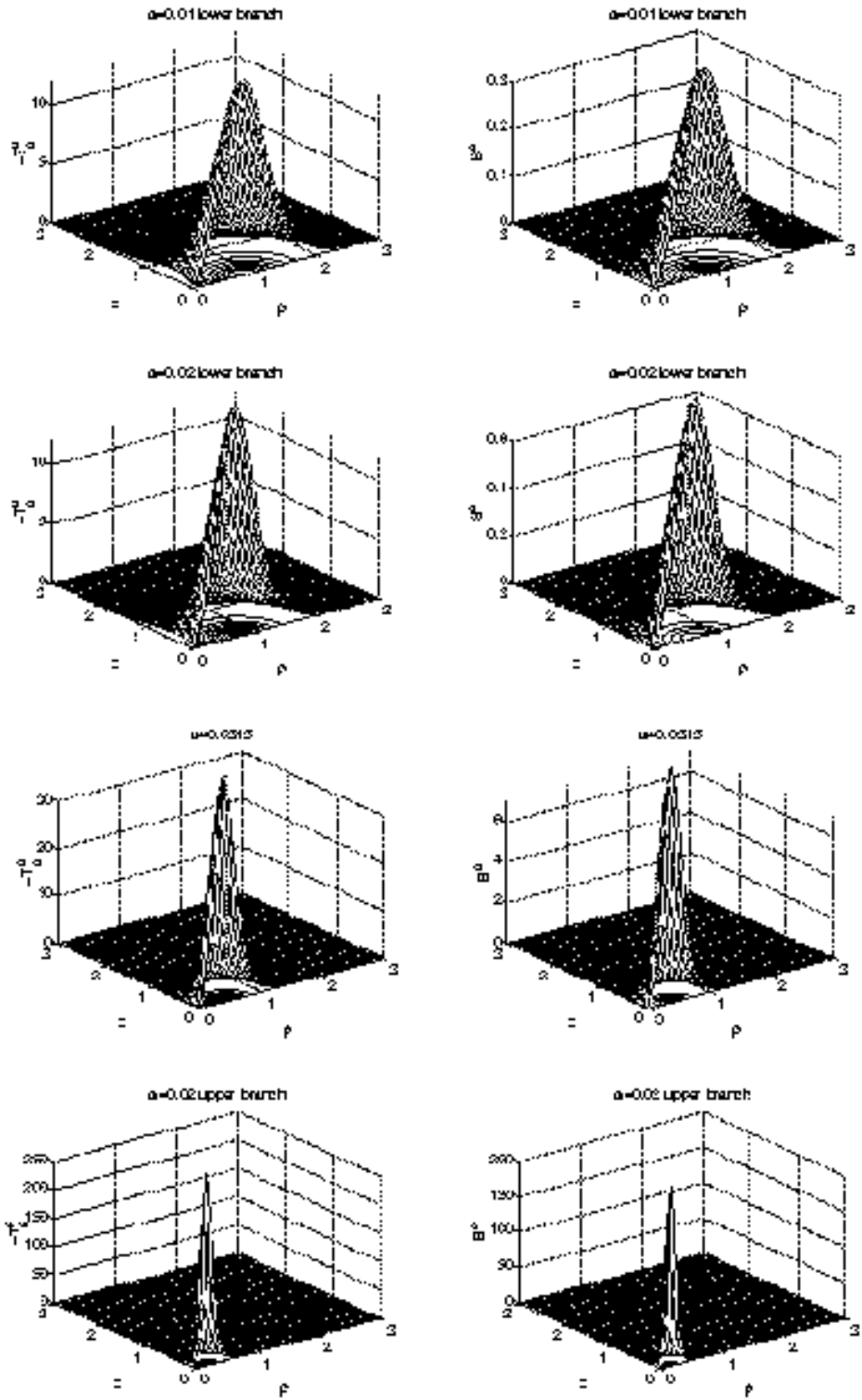}}
\vspace{-4.cm}
}}
\caption{\label{Fig2}The energy density $\epsilon = T_0^0$ (left) and the 
baryon number density $B^0$ (right) are shown for the $SU(2)$ $B=2$ solution. }
\end{figure}

\begin{figure}
\parbox{\textwidth}
{\centerline{
\mbox{
\epsfysize=25.0cm
\epsffile{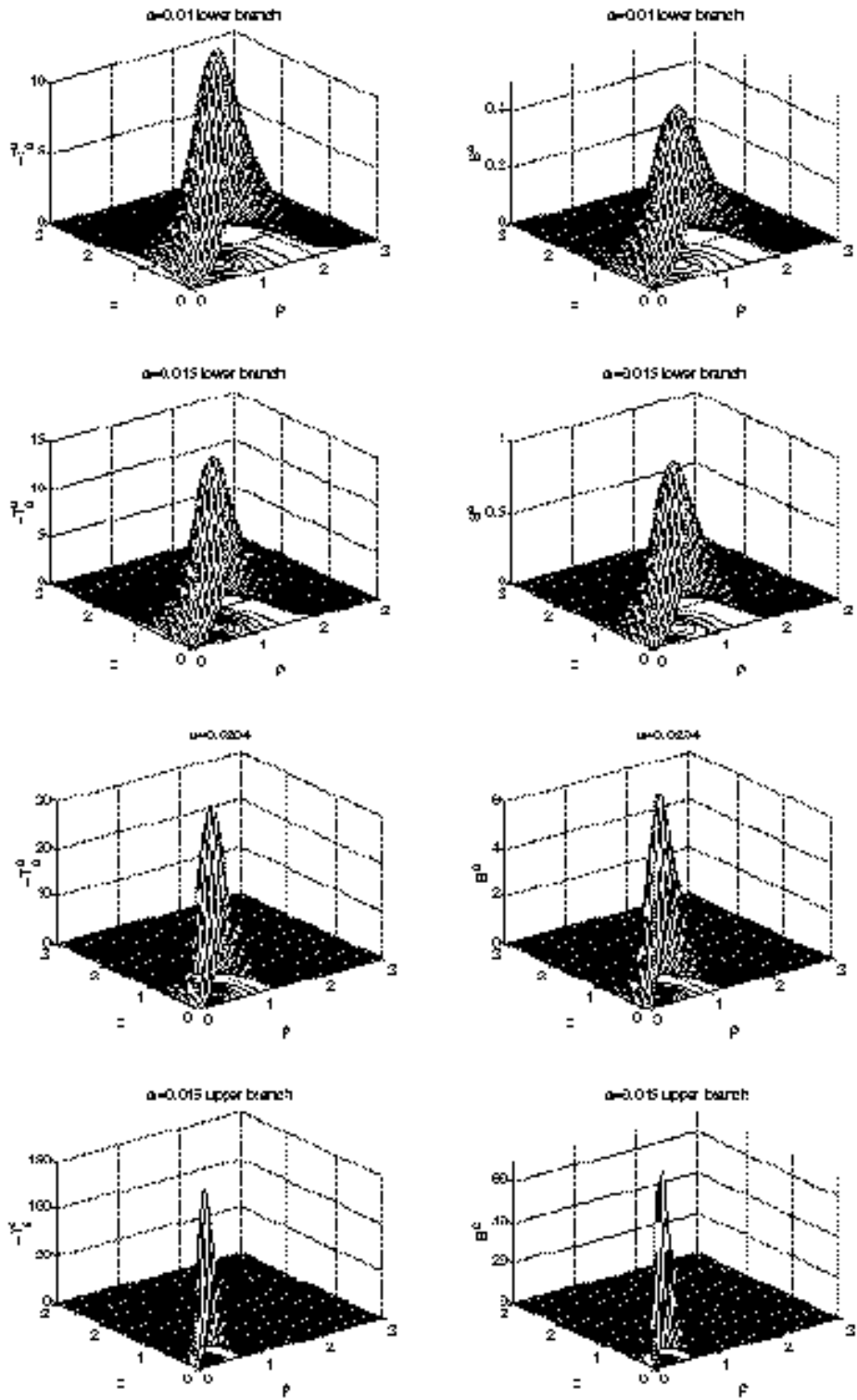}}
\vspace{-4.cm}
}}
\caption{\label{Fig3} Similar as Figure 3 for the SU(3) $B=4$  solution.}
\end{figure}

Figure \ref{Fig2} and \ref{Fig3} present the dimensionless energy density 
$-T_0^0$ and the 
baryon density $B^0$
as functions of $\rho=x\sin\theta$ and $z=x\cos\theta$ for several values 
of the 
coupling constant $\al$ for the SU(2) $B=2$ and SU(3) $B=4$ 
skyrmions.
In the  SU(2) $B=2$ case, both
$T_0^0$ and $B^0$ possess a  maximum on the $\rho$-axis which corresponds
to a ring in the  $xy$-plane; in accordance with the toroidal symmetry of
the  SU(2) $B=2$ non-gravitating skyrmion.
The radius of the torus decreases with increasing $\alpha$
on the lower branch. On the upper branch it continues to decrease 
with decreasing $\alpha$ and tends to zero as $\alpha$ tends to zero.
The height of the maximum increases on the lower branch with increasing 
$\alpha$ and on the upper branch with decreasing $\alpha$  and 
diverges on the upper branch as $\alpha$ tends to zero.
On the other hand, the SU(3) $B=4$  gravitating skyrmion has the shape of 
a double torus  as can be observed 
directly from  Figure \ref{Fig3} (similar to the flat case).
Note  that, as $\alpha$ increases the radii of the tori and their distance
from the $xy$-plane decrease along the 
lower branch and shrink to zero along the upper branch as
$\alpha$ decreases. Simultaneously, the maxima of $T_0^0$ and $B^0$ increase 
and diverge on the upper branch as $\alpha$ tends to zero.
This singular behaviour is seen in our choice of the dimensionless
radial coordinate $x=e\kappa r$ as first observed by Bizon and Chmaj \cite{BZ}.
If the rescaled coordinate  $\tilde{x}=x/\sqrt{\alpha}$ is used instead, 
all functions are smooth and the rescaled energy 
$\tilde{M}/B= \sqrt{\alpha} M/B$
remains finite as $\alpha$ tends to zero on the upper branch.

Finally we note that, for the SU(2) $B=2$ and 
SU(3) $B=4$ gravitating skyrmions 
the dependence on the coupling constant $\alpha$ follows the same pattern
as for the spherically symmetric SU(2) and SU(3) skyrmions.

\section{Conclusions}

In this paper, we have used the improved  harmonic map
ansatz to look at field configurations involving gravitating skyrmions.
In particular we have derived approximations of the  $SU(2)$  $B=2$ and $SU(3)$
 $B=4$ gravitating skyrmions. 
Our configurations are not solutions of the equations
of motion but we believe that they are close to them; hence we hope 
that our results 
are close to what would have been seen 
for the solutions.

We have found that, 
(qualitatively) the situation is very similar to the 
 four spherical symmetric $SU(3)$ skyrmions \cite{BHIZ} since 
 the equations  have solutions only for a range of the gravitational 
coupling constant with the skyrmions being bound by the gravitational field. 
That is,  for  values
of the coupling constant below its critical value - we have two solutions 
(of which one has much higher energy) and when the coupling constant 
goes beyond its critical value - there are not unitary solutions for the Skyrme
 fields suggesting that  the system possesses two complex solutions. 
The shapes of the energy densities 
are very similar to what is seen for the non-gravitating skyrmions. 
Thus we suspect
that the effects of the gravitational field are unversal in nature; 
the field binds
the skyrmions and it alters their properties in a universal way making them 
more compact but not changing their basic shape.

\end{document}